\documentclass{iopjournal}

\usepackage{graphicx}
\usepackage{hyperref}
\usepackage{amsmath}
\usepackage{courier}
\usepackage{placeins}
\usepackage{setspace}
\usepackage{listings}
\usepackage{listings}
\usepackage{xcolor}
\usepackage{float}
\usepackage{matlab-prettifier}
\usepackage{pythonhighlight}
\usepackage{multirow}
\usepackage{undertilde}
\usepackage{natbib}
\usepackage{etoolbox}

\hypersetup{colorlinks=true, citecolor=blue, urlcolor=blue, linkcolor=blue}
\graphicspath{{./Downloads/}}
\makeatletter

\patchcmd{\NAT@citex}
  {\@citea\NAT@hyper@{%
     \NAT@nmfmt{\NAT@nm}%
     \hyper@natlinkbreak{\NAT@aysep\NAT@spacechar}{\@citeb\@extra@b@citeb}%
     \NAT@date}}
  {\@citea\NAT@nmfmt{\NAT@nm}%
   \NAT@aysep\NAT@spacechar\NAT@hyper@{\NAT@date}}{}{}

\patchcmd{\NAT@citex}
  {\@citea\NAT@hyper@{%
     \NAT@nmfmt{\NAT@nm}%
     \hyper@natlinkbreak{\NAT@spacechar\NAT@@open\if*#1*\else#1\NAT@spacechar\fi}%
       {\@citeb\@extra@b@citeb}%
     \NAT@date}}
  {\@citea\NAT@nmfmt{\NAT@nm}%
   \NAT@spacechar\NAT@@open\if*#1*\else#1\NAT@spacechar\fi\NAT@hyper@{\NAT@date}}
  {}{}

\makeatother

\begin{document}
\articletype{Paper}

\title{Noise Propagation and Non-Uniqueness in Three-Material Dual Energy CT with a Path Length Constraint}
\date{2026}
\author{Andrew E Ferretti$^{*1}$\orcid{0009-0001-0771-3084}, Emil Y Sidky$^1$\orcid{0000-0002-6951-2456}, JP J Phillips$^1$\orcid{0000-0002-3437-425X}, Xiaochuan Pan$^1$\orcid{0000-0002-3074-9771}}

\affil{$^1$Department of Radiology, The University of Chicago, Chicago, IL 60637, United States of America}

\affil{$^*$Author to whom any correspondence should be addressed.}

\email{aeferretti@uchicago.edu}


\date{\today}

\begin{abstract}
Objective: Using a path length constraint with dual energy computed tomography (DECT) data, a sample can be decomposed into three basis materials. We aim to analyze the stability of such a system by presenting a noise propagation framework for optimizing the contrast to noise ratio (CNR) of virtual monochromatic images (VMIs) and by studying instances where sinogram decompositions are not unique.
Approach: We extend a noise propagation framework developed for two-material DECT to three materials. Using the Jacobian of the system, we identify tube voltage pairings with good noise properties and obtain analytical expressions for VMI CNR. We apply spectral theorem to identify the best CNR possible in a linear combination of basis images. By using a reduced two-dimensional problem, we develop the sampling isoline recovery algorithm (SIRA) to study non-unique sinogram decompositions in this system. 
Main results: Using our noise propagation framework with a realistic breast phantom consisting of adipose, glandular tissue, and iodine, we find an ideal tube pairing of 34 kVp and 69 kVp, and a maximum VMI CNR of a spiculated lesion of 9.9. By allowing arbitrary linear combinations of basis material images, we find a CNR of 10.2 can be achieved. In a realistic phantom with a cardiac plaque, we find non-unique water, bone and iodine decompositions, and show that qualitatively accurate reconstructions can be obtained using SIRA.
Significance: We developed a framework for direct optimization of VMI CNR in a three-material DECT decomposition and showed that an arbitrary linear combination of basis material images can achieve a higher CNR than a strict VMI. We demonstrate the presence of non-unique sinogram decompositions in three-material DECT with a data constraint. 
\end{abstract}

\section{Introduction}

Dual energy computed tomography (DECT) collects X-ray projections with two distinct X-ray energy spectra, which in turn allows the distribution of two materials in a sample to be estimated. The idea of DECT was first conceptualized by \citet{hounsfield_computerized_1973}, and a full mathematical treatment of DECT was supplied by \citet{r_e_alvarez_energy-selective_1976}. DECT relies on an assumption that the energy-dependent linear attenuation coefficients (LACs) of the object being imaged lie in a two-dimensional space (e.g., energy-dependent Compton and photoelectric terms, or LACs from two distinct materials) so that two spectral measurements are sufficient to recover all information about the LACs of the object. For soft tissues, an LAC dimensionality of two is a good assumption, as shown using singular-value decomposition techniques by \citet{alvarez_7_1986} and \citet{tang_conditioning_2021}. However, when higher Z materials are introduced, notably with iodine contrast, this is no longer the case, with some estimates in the literature putting the dimensionality of the LAC space as high as four \citep{tang_conditioning_2021,alvarez_energy_1982, bornefalk_xcom_2012, sukovle_basis_1999}. This opens the possibility of material decomposition in DECT with three or more materials. A number of potential applications have been explored with three-material decompositions in DECT, including cardiac plaque detection \citep{boussel_photon_2014, gronberg_feasibility_2020}, bone edema detection \citep{liu_dual-energy_2022}, bone mineral density assessment \citep{goodsitt_two_1987}, liver fat assessment \citep{mendonca_flexible_2014, fischer_quantification_2011} and imaging with dual contrast agents \citep{ren_dual-contrast_2022, ataei_visual_2024}.

With only two spectral measurements, three-material decomposition by itself is not directly possible. In practice three-material decomposition is achieved with a third spectral measurement, which might be provided by a photon counting system, or by adding an additional data constraint. Note that material decomposition is usually performed either in projection space or image space, each of which lend themselves to different data constraints. In projection space, it is possible to use a path length constraint to perform three-material decompositions, first used by \citet{kis_single-energy_2012} for two-material decompositions from single-energy data. Alternative approaches in image space have used variations of a volume constraint \citep{yu_pre-reconstruction_2009, liu_model-based_2022, long_multi-material_2014}, including a robust method by \citet{mendonca_flexible_2014}, or a mass fraction constraint \citep{liu_quantitative_2009}. There have also been efforts to use convolutional neural networks to obtain three-material decompositions \citep{zhu_feasibility_2022, gong_deep-learning-based_2020}. 

When presenting DECT images, it is common to combine material decompositions into a single virtual monochromatic image (VMI). For two-material decompositions, the behavior of the contrast-to-noise ratio (CNR) of VMIs has been studied in a myriad of approaches. \citet{yu_image_2009, yu_virtual_2011} explore how patient size and radiation dose partitioning impacts CNR in VMIs produced in the image domain. For projection-based decompositions, expressions for the variances of DECT material basis sinogram elements have been derived \citep{r_e_alvarez_energy-selective_1976, alvarez_comparison_1979}, used to estimate pixel variances in material decompositions with a filtered backprojection reconstruction \citep{alvarez_noise_1976}, and likewise applied to predict CNR as a function of VMI energy for a simple geometric phantom \citep{terzioglu_optimizing_2024}. More empirical approaches exploring dependencies of CNR on tube energy have also been taken, such as those by \citet{michalak_selection_2017} and \citet{dabli_optimization_2021}. For three-material decompositions, the Cramer-Rao lower bound has been employed to study the variance of material decomposition sinograms for multi-energy CT (MECT) \citep{alvarez_dimensionality_2013,ding_invertibility_2021}. \citet{ren_three-material_2018, ren_impact_2019} demonstrated that the use of prior information in DECT or MECT can improve the noise performance of these systems. Although there has been much work on reducing noise and optimizing VMI CNR in two-material decompositions and some parrallel efforts for three-material decompositions, direct CNR optimization in VMIs formed from three-material decompositions using data constraints in DECT is relatively unexplored. 

It has also been demonstrated that there are cases in DECT where material decompositions are not unique, i.e., where due to the non-linearity of the DECT problem, two sets of material path lengths satisfy the same DECT data \citep{levine_nonuniqueness_2017, phillips_non-unique_2024}. In response, there has been an effort to establish more sufficiency conditions for unique material decomposition for DECT and MECT \citep{alvarez_invertibility_2019, bal_uniqueness_2020, ding_invertibility_2021}. These approaches tend to rely on calculating the Jacobian of the material sinogram to DECT measurement transform, for example using P-matrices \citep{gale_jacobian_1965} or the Hadamard global inverse theorem. While these analyses translate to three-material DECT achieved with different constraint conditions, the invertibility and uniqueness of DECT three-material decompositions using a data constraint has not been studied directly. 

In this work, we aim to use knowledge of the material sinogram to DECT measurement transform Jacobian to study the noise and uniqueness properties of dual energy three-material decomposition with a path length constraint. First, by expanding the noise propagation approach of \citet{terzioglu_optimizing_2024}, we derive analytic expressions for sinogram and VMI variances as a function of tube energy and VMI energy. We then use these expressions to show how an optimal tube energy can be selected for a realistic uncompressed breast phantom, and present theory-based CNR calculations as a function of VMI energy to optimize the choice of weight when combining material basis images into a VMI. Second, we use the Jacobian with a realistic cardiac phantom with a calcified plaque to find tube pairings where injectivity is not guaranteed. We develop the sampling isoline recovery algorithm (SIRA) based on the work of \citet{phillips_dual-energy_2025} for finding non-unique solutions in standard DECT and present reconstructions where non-unique material decompositions are present. Crucially, our approach relies on a reduced two-dimensional problem, where the sinogram estimation problem is equivalent to finding the intersections of two level curves.

\section{Methods}
\subsection{Data Models and Inversion} 
For an $n$-material basis image assumption, we approximate the linear attenuation coefficient as 
\begin{equation}
\mu(E,\ y) \approx \sum_{j=1}^{n} M_j(E)f_j(y),
\label{mu_eq}
\end{equation}
where $M_j(E)$ are the linear attenuation coefficients of the $n$ materials at a particular energy $E$, and $f_j(y)$ are the corresponding material volume fractions at position $y$ \citep{r_e_alvarez_energy-selective_1976}. For this work, we typically set $n=3$ to study a three-material system. The material path lengths of the $l$th ray of the $j$th material can be written as 
\begin{equation}
x_{l,j} = \int_l f_j(y)dy.
\end{equation}
Then for a given tube potential, the expected X-ray intensity along a ray will be given by  
\begin{equation}
\phi_i(\vec{x}_l) = \int_0^{V_i} S_i(E)D(E)e^{- \vec{x}_l\cdot\vec{M}(E) }dE, 
\label{int_eq}
\end{equation} 
where $\vec{x}_l$ is a three-element column vector representing the material line integrals of the basis materials $x_{l,j}$, $\vec{M}(E)$ is a three-element column vector of the energy-dependent linear attenuation coefficients $M_j(E)$, $S_i(E)$ represents the X-ray spectrum produced for the $i$th accelerating tube potential $V_i$, and $D(E)$ models a detector response taken to be linear ($D(E)=\alpha E$). We work with X-ray data after the application of a negative log operation,
\begin{equation}
g_i(\vec{x}_l) = -\ln\left(\int_0^{V_i} S_i(E)D(E)e^{- \vec{x}_l\cdot\vec{M}(E) }dE\right) = -\ln(\phi_i(\vec{x}_l)),
\label{meas_eq}
\end{equation}
where $g_i$ is referred to as our measurement. 

Specifically, we study a three-material system using X-ray measurements with two distinct spectra ($S_i(E)$) and a path length constraint, which generally must be known from some \textit{a priori} means, like prior CT measurements. The path length constraint under a three-material assumption can be stated as 
\begin{equation}
d_l=x_{l,1} + x_{l,2} + x_{l,3},
\label{dist_con_eq}
\end{equation}
which with equation \ref{meas_eq} for $i$=1, 2, comprises a non-linear system of three equations and three unknowns. Image reconstruction for this problem can now be characterized as a two-step process. First, the sinogram recovery step produces three basis sinograms by solving the non-linear system of equations

\begin{align}
	g_1(\vec{x}_l) &= -\ln\left(\int_0^{V_i} S_i(E)D(E)e^{- \vec{x}_l\cdot\vec{M}(E) }dE\right) \label{data_eq1} \\
	g_2(\vec{x}_l) &= -\ln\left(\int_0^{V_i} S_i(E)D(E)e^{- \vec{x}_l\cdot\vec{M}(E) }dE\right) \label{data_eq2}\\
	d_l&=x_{l,1} + x_{l,2} + x_{l,3} \label{data_eq3}, 
\end{align}
which can be done numerically using algorithms like Newton's method. Second, filtered backprojection (FBP) of each material sinogram is performed to produce a set of material basis images. 

\subsection{Noise Models}
We aim to model the statistical properties of the pixels in an image reconstructed from an aggregated basis sinogram formed from the basis sinograms as
\begin{equation}
x^{agg}_l = w_1x_{l,1} + w_2x_{l,2} + w_3x_{l,3} = \vec{w}\cdot\vec{x}_l, 
\label{agg_eq}
\end{equation}
where the weights $w_j$ form an arbitrary unit vector. We call such an image an aggregated basis image (ABI). Note that the weights in equation \ref{agg_eq} are almost always chosen using normalized linear attenuation coefficients of the basis materials $w_j=M_j(E)/||\vec{M}(E)||_2$ at a given energy, and that under such a choice, the reconstructed image is called a virtual monochromatic image (VMI) and can be considered a special case of an ABI. 

All three measurements can then be modeled as random variables with known probability distributions. The X-ray measurements for the $l$th ray $\boldsymbol{g}_{l,1}^M$ and $\boldsymbol{g}^M_{l,2}$ (where boldface denotes a random variable) can be modeled as compound Poisson random variables, where the number of photons at each energy can be considered a separate Poisson random variable, and the integrated response of the detector is their weighted sum \citep{whiting_signal_2002}. For example, for tube potential $V_i$, the original intensity data measured by the detector can be approximated for a discretized energy spectra with $N$ energy bins as 
\begin{equation}
\boldsymbol{\phi}_i^M = \sum_{k=1}^{N}D(E_k) \boldsymbol{\lambda}_i(E_k)\Delta E,
\end{equation}
where $\boldsymbol{\lambda}_i(E)$ is a Poisson random variable with mean $S_i(E)e^{-\vec{M}(E)\cdot \vec{x_l}}$. For a continuous spectrum, the variance of the intensity measurement has been shown to be 
\begin{equation}
\sigma_{\phi,i}^2 =  \int_0^{V_i} S_i(E)D^2(E) e^{- (\vec{x}_l\cdot\vec{M}(E))}dE,
\label{data_std_eq}
\end{equation}
which in turn can be used to obtain the variance of $\boldsymbol{g}^M_{l,i}$ using the simple relation that for a function of a random variable $\boldsymbol{y}=f(\boldsymbol{x})$, 
\begin{equation}
\sigma_y^2\approx( f'(\bar{x}))^2\sigma_x^2,
\end{equation}
where $\bar{x}$ and $\sigma_x^2$ and are the mean and variance of a random variable. Because the means of $\boldsymbol{g}^M_{l,1}$ and $\boldsymbol{g}^M_{l,2}$, $\bar{g}^M_{l,1}$ and $\bar{g}^M_{l,2}$, are given by equation \ref{meas_eq}, altogether the mean and variance of the X-ray measurements can be characterized as 

\begin{align}
\bar{g}^M_{l,i} &= -\ln\left(  \begingroup\textstyle\int\endgroup_0^{V_i} S_i(E)D(E)e^{- \vec{x}_l\cdot\vec{M}(E) }dE\right)  \label{real_x_mean} \\
\sigma_{l,i}^2 &= \frac {\int_0^{V_i} S_i(E)D^2(E) e^{- (\vec{x}_l\cdot\vec{M}(E))}dE} {\left( \int_0^{V_i} S_i(E)D(E)e^{- \vec{x}_l\cdot\vec{M}(E) }dE  )\right)^2} \ \ \ .
\label{mean_and_variance_eq}
\end{align}
The statistical uncertainty of the path length measurement depends on how it is obtained. Because many natural processes are well-approximated by a Gaussian distribution, we model the path length measurement $\boldsymbol{d}^M_l$ as a Gaussian random variable with mean equal to $\bar{d}_l$ and a constant variance of $\sigma_{d}$. For example, if $\boldsymbol{d}^M_l$ is obtained from past CT reconstructions, $\sigma_{d}$ might be taken as one pixel width.

\subsection{Propagation of the Mean}
The expected values of $\boldsymbol{g}^M_1$, $\boldsymbol{g}^M_2$, and $\boldsymbol{d}^M_l$ can be propagated under a series of linear approximations to obtain the mean pixel values of an ABI. For a given phantom, the true values of $\vec{x}_l$ are computed using any forward projection algorithm (e.g., the Siddon ray-tracing algorithm \citep{siddon_fast_1985}). The means of $\boldsymbol{g}^M_1$ and $\boldsymbol{g}^M_2$ are then obtained using equation \ref{real_x_mean}, while the mean of $\boldsymbol{d}^M_l$ for each sinogram element is obtained from equation \ref{dist_con_eq}. Taking a first-order Taylor expansion of the expectation values of $\boldsymbol{g}^M_1$, $\boldsymbol{g}^M_2$, and $\boldsymbol{d}_l$ about the means of the basis sinogram elements, $\bar{x}_{l,1}$, $\bar{x}_{l,2}$, and $\bar{x}_{l,3}$, then gives
\begin{align}
       \bar{g}^M_{l,1}&= g_1(\bar{x}_{l,1},\ \bar{x}_{l,2},\ \bar{x}_{l,3}) \label{mean1}\\
       \bar{g}^M_{l,2}&= g_2(\bar{x}_{l,1},\ \bar{x}_{l,2},\ \bar{x}_{l,3}) \label{mean2}\\
       \bar{d}^M_l &= \bar{x}_{l,1} + \bar{x}_{l,2} + \bar{x}_{l,3} \label{mean3}. 
\end{align}
This suggests that a good approximation of $\bar{x}_{l,1}$, $\bar{x}_{l,2}$, and $\bar{x}_{l,3}$ can be obtained by numerically solving the non-linear system of equations represented in equation \ref{mean1}-\ref{mean3}. 

However, this system is not always well-behaved, such that we require that the Jacobian of the basis sinogram to measurement transform $J(\vec{x}_l)$ have a non-zero determinant over all possible basis sinogram values $\vec{x}_l$ of interest, where $J(\vec{x}_l)$ can be written as
\begin{equation}
J(\vec{x}_l) =\begin{bmatrix} \frac{\partial g_1}{\partial x_{l,1}} & \frac{\partial g_1}{\partial x_{l,2}}  & \frac{\partial g_1}{\partial x_{l,3}} \\ \frac{\partial g_2}{\partial x_{l,1}} & \frac{\partial g_2}{\partial x_{l,2}}  & \frac{\partial g_2}{\partial x_{l,3}} \\ 1& 1 & 1 \end{bmatrix},
\label{jacques}
\end{equation}
with evaluations at $\vec{x}_l$ suppressed for notational convenience. To calculate $J(\vec{x}_l)$, it is helpful to introduce the quanity $\mathcal{I}$, where we define 
\begin{equation}
\mathcal{I}_{i,j}(\vec{x}_l) = \int_0^{V_i} S_i(E)D(E)M_j(E)e^{-\vec{x}_l\cdot\vec{M}(E) }dE,\ i=1,\ 2,\ j=1,\ 2,\,3. 
\label{i_note}
\end{equation}
The partial derivatives of the measurements with respect to the path lengths of the basis materials are then given by 
\begin{equation}
\frac{\partial g_i(\vec{x}_l)}{\partial x_{l,j}} = \frac{\mathcal{I}_{i,j}(\vec{x}_l)} {\phi_i(\vec{x}_l)},\ i=1,\ 2,\ j=1,\ 2,\ 3, 
\label{deriv_eq}
\end{equation}
recalling that $\phi_i(\vec{x}_l)$ is the X-ray intensity along a ray. Note that under the condition of det$|J(\vec{x}_l)|$ not equal to zero and strictly three unique materials in the phantom, the step of estimating $\bar{x}_{l,1}$, $\bar{x}_{l,2}$, and $\bar{x}_{l,3}$ is essentially trivial (i.e., $\bar{x}_{l,1}=x_{l,1}$ within machine precision).

The mean of the aggregated sinogram $\bar{x}^{agg}_l$ for a particular weight $\vec{w}$ can then be obtained from $\bar{x}_{l,1}$, $\bar{x}_{l,2}$, and $\bar{x}_{l,3}$ using equation \ref{agg_eq} under a similar first-order Taylor expansion. Likewise, the mean of the ABI pixels is calculated from $\bar{x}^{agg}_{l}$ using FBP. It is useful to write this explicitly using the discretized FBP matrix as
\begin{equation}
\bar{y}^{agg} = B \bar{x}^{agg},
\end{equation}
where $\bar{y}^{agg}$ is a vector representing the reconstructed image matrix, and $B=(b_{pl})_{1<p<P, 1<l<L}$ is the FBP matrix for a sinogram of $L$ elements and image matrix of $P$ pixels.

\subsection{Propagation of the Variance}
We outline the propagation of the variances of $\boldsymbol{g}^M_{l,1}$, $\boldsymbol{g}^M_{l,2}$, and $\boldsymbol{d}^M_l$ to obtain ABI pixel variances by extending a noise propagation framework for a two-material system to a three-material system with a path length constraint relying on the Jacobian of the material sinogram to measurement transform as first described by \citet{terzioglu_optimizing_2024}.

Assuming independence between the three measurements, the covariance matrix of the measurements is just 
\begin{equation}
\Sigma^{meas}_l =\begin{bmatrix} \sigma^2_{l,1} & 0  & 0 \\ 0 & \sigma^2_{l,2} & 0 \\ 0 & 0 & \sigma_{d}^2\end{bmatrix},
\end{equation}
where $l$ indexes measurement along a particular ray, $\sigma_{l,i}$ can be obtained from equation \ref{mean_and_variance_eq}, and  $\sigma_{d}$ is an assumed constant. From the measurement covariances, a first-order Taylor expansion gives the covariance matrix of the basis sinogram elements as 

\begin{equation}
\Sigma^{basis}_l= J^{-1}(\vec{\bar{x}}_l)\Sigma^{meas}_l(J^{-1}(\vec{\bar{x}}_l))^{\top},
\label{jac_covar}
\end{equation}
where the $3\times3$ matrix $J^{-1}(\vec{\bar{x}}_l)$ is the inverse of the material sinogram to measurement transform Jacobian as in equation \ref{jacques} evaluated at $\bar{x}_{l,1}$, $\bar{x}_{l,2}$, and $\bar{x}_{l,3}$ \citep{cowan_statistical_1998}. Note that by ignoring higher-order terms, we implicitly model $\boldsymbol{g}^M_{l,1}$, $\boldsymbol{g}^M_{l,2}$, and $\boldsymbol{d}^M_l$ as being Gaussian and assume they are completely characterized by a mean and standard deviation. The variances of the aggregated sinogram elements as calculated from equation \ref{agg_eq} will then be 
\begin{equation}
(\sigma^{agg}_l)^2 = \vec{w}^{\top}\Sigma^{basis}_l\vec{w}.
\end{equation} 
Making the assumption that each line integral in the aggregated basis sinogram is independent for different rays, their covariances $\Sigma^{agg}$ can be described by an $L\times L$ diagonal matrix with elements $(\sigma^{agg}_l)^2$ on its diagonal. The covariance matrix of the final pixels in the ABI is then
\begin{equation}
\Sigma_y = B\Sigma^{agg} B^{\top},
\end{equation} 
where $B$ is again the FBP matrix.

\subsection{Contrast-to-Noise Ratio Calculation and Optimization}
For some signal region, we define the contrast-to-noise ratio (CNR) of an ABI as
\begin{equation}
CNR =  \left( \frac{1}{N}\sum_{n=1}^N \bar{y}^{agg}_n - \frac{1}{M}\sum_{m=1}^M \bar{y}^{agg}_m\right)/s_{signal},
\label{cnr_eq}
\end{equation}
where $\bar{y}^{agg}_n$ indexes one of $N$ pixels in the signal region, $\bar{y}^{agg}_m$ indexes one of $M$ pixels in the background region, and $s^2_{signal}$ is the mean of the variances of pixels in the signal region. For a given signal region, $s^2_{signal}$ can always be obtained by averaging over the appropriate elements of $\Sigma_y$. Alternatively, we can use $\Sigma^{basis}_l$ to define a $3\times3$ matrix $Q$ whose diagonal elements are the mean pixel variances of each material density map for some signal region and whose off-diagonal elements are the mean pixel covariances between them in the same signal region. It has been shown by \citet{terzioglu_optimizing_2024} that entries of $Q$ are given by 
\begin{equation}
q_{ij} = \frac{1}{P}\sum_{l=1}^{L}\sum_{p=1}^{P}b^2_{pl}\cdot(\Sigma^{basis}_l)_{ij}.
\label{quelements_eq}
\end{equation}
Moreover, the mean pixel variance of the ABI is related to the matrix $Q$ and is given by 
\begin{equation}
s_{signal}^2 = \vec{w}^TQ\vec{w},
\label{s2_eq}
\end{equation}
providing a computationally efficient means of calculating $s_{signal}^2$ for different choices of ABI weight $\vec{w}$. Analytically calculating the CNR as in equation \ref{cnr_eq} for a particular choice of $\vec{w}$ then amounts to finding the matrix $Q$ for a given phantom and evaluating the mean of the signal and background regions from the calculated pixel means. 

Furthermore, a choice of arbitrary weights $\vec{w}$ that minimizes $s_{signal}^2$ can be selected using the matrix $Q$. In particular, because the matrix $\Sigma^{meas}_l$ is symmetric and diagonal, if det$|J(\vec{x}_l)| \neq0$ it follows that $\Sigma^{basis}_l$ is a positive definite matrix, as is $Q$. By spectral theorem, $Q$ is then diagonalizable, so that $Q\vec{u}_i=\lambda_i\vec{u}_i$ for $i=1,\ 2,\ 3$, where $\vec{u}_i$ is a unit eigenvector of $Q$, $\lambda_i$ is its associated eigenvalue, and $\lambda_1 >\lambda_2 > \lambda_3 > 0$. Then minimizing $s_{signal}^2$ becomes 
\begin{equation}
\underset{||w||_2}{\text{min}}(s^2(V_1,\ V_2,\ \vec{w})) = \underset{||w||_2}{\text{min}}( \vec{w}^{\top}Q(V_1,\ V_2)\vec{w}) = \lambda_3(V_1,\ V_2)
\end{equation}
where the minimizer at a particular tube pairing is $\vec{w}=\vec{u}_3(V_1,\ V_2)$. Generally minimizing $s_{signal}^2$ is a matter of finding the minimum $\lambda_3(V_1,\ V_2)$ over all tube potential pairings. As analogously noted by \citet{terzioglu_optimizing_2024}, equations \ref{jac_covar} and \ref{quelements_eq} suggest that both the sinogram covariances and the elements of $Q$ are inversely proportional to the determinant of the the material sinogram to measurement transform Jacobian. Because the smallest diagonal entry of a matrix provides an upper bound for it's smallest eigenvalue, a tube pairing that maximizes det$|J(\vec{x_l})|$ should minimize the smallest eigenvalue of $Q$, and therefore also reduce $s_{signal}^2$ \citep{horn_matrix_2017}.

\subsection{Sampling Isoline Recovery Algorithm for Non-Unique Solutions}
In general, because the material sinogram to measurement transform is non-linear, a decomposition into basis sinograms may not be unique. That is, there may not be an unique combination of basis sinograms $x_{l,1}$, $x_{l,2}$, and $x_{l,3}$ that correspond to a particular set measurements, $g^M_1$, $g^M_2$, and $d^M_l$. This might occur in cases where det$| J(\vec{x}_l)|=0$, for example. Tube pairings where unique solutions of the recovered basis sinograms do not exist may still find practical use. Ideally, in such cases the sinogram recovery step should return all possible path length solutions. We describe the sampling isoline recovery algorithm (SIRA) for finding multiple solutions for this three-material problem using a reduced system of equations. 

First, to simplify the form of the resulting two-dimensional problem, it is useful to define new effective linear attenuation coefficients for two basis materials $M_1'(E) = M_1(E)-M_3(E)$ and $M_2'(E)=M_2(E)-M_3(E)$, and a new spectral term $A_l(E)$ dependent on $d_l$ and $M_3(E)$ where $A_l(E)=e^{-d_l\cdot M_3(E)}$. Equations \ref{data_eq1}-\ref{data_eq3} can then be re-formulated as a reduced system of two non-linear equations $g_i^R$,
\begin{align}
	g^R_1(x_{l,1},\ x_{l,2}) = -\ln\left(\int_0^{V_1} S_1(E)D(E)A_l(E)e^{-M'_1(E)x_{l,1} - M'_2(E)x_{l,2}}dE\right) &  \label{reduced_sys_eq1}\\
	g^R_2(x_{l,1},\ x_{l,2}) = -\ln\left(\int_0^{V_2} S_2(E)D(E)A_l(E)e^{-M'_1(E)x_{l,1} - M'_2(E)x_{l,2}}dE\right) &. \label{reduced_sys_eq2}
\end{align}
We then consider the implicit form of equations \ref{reduced_sys_eq1} and \ref{reduced_sys_eq2} for a particular path length and measured X-ray data $g_i^M$,
\begin{equation}
g_i^R(x_1,\ x_2) - g_i^M = 0,
\label{level_set_eq}
\end{equation}
where we drop the ray-dependent indication $l$ on the material path lengths $x_1$ and $x_2$ for notational convenience. Crucially, we observe that the material path lengths that satisfy equation \ref{level_set_eq} for a given spectra are simply a level curve of the function $g_i^R(x_1,\ x_2)$ at level $g_i^M$. Finding all material path lengths solutions $(x_1,\ x_2)$ that could lead to the observed data then amounts to finding the intersection of these level curves for two levels $g_1^M$ and $g_2^M$. Note that given a $(x_1,\ x_2)$ pair, the path length of the third material can be obtained using equation \ref{dist_con_eq}. 

Then to find non-unique solutions, we need to determine the level curves of $g_i^R(x_1\ ,x_2)$ in some domain of path lengths of the first basis material, $D_{x_1}=\{x_1^{(min)},\ x_1^{(max)}\}$. To calculate level curves in practice, any single level curve of $g_i^R(x_1,\ x_2)$ can also be considered as an implicit function, i.e., $x_2$ = $\tau_i(x_1)$, where $\tau_i(x_1)$ is a function of the path length of basis material one that returns the path length of basis material two $x_2$ that satisfies $g_i^R(x_1,\ x_2) - g_i^M = 0$ given a fixed input $x_1$. Note that $\tau_i(x_1)$ is also effectively a function of the input measurement $g_i^M$, so that a unique implicit function $\tau_i(x_1)$ exists for each sinogram element. While an analytic form of $\tau_i(x_1)$ is difficult to obtain, it can be evaluated numerically for a given $g_i^M$ by finding the root of $g_i^R(x_1,\ x_2) - g_i^M$ for a chosen path length of material one in the domain $D_{x_1}$. In particular, a Newton's method routine is used with an initial guess of $(x_2)_0$, for which pseudo-code is developed in Appendix \ref{psuedo} in Algorithm \ref{algo_2}.

Given a method for evaluating level curves $\tau_i(x_1)$ for the two energy spectra $V_i$, determining solutions to the system of equations then amounts to finding where $\tau_1(x_1) = \tau_2(x_1)$, or equivalently, finding roots of $\tau_1(x_1) - \tau_2(x_1)$. The isoline recovery method of \citet{phillips_dual-energy_2025} approached this problem by determining the number of solutions based on the sign of $\tau_1(x_1) - \tau_2(x_1)$ at the edges of the domain $D_{x_1}$ and the location of an assumed single stationary point to sub-divide the domain into monotonic regions to root-find in (where root-finding methods like bisection are guaranteed to find a root). Empirically,  $\tau_1(x_1) - \tau_2(x_1)$ can have multiple stationary and inflection points, so that such an approach is no longer feasible. Instead, we sample the curve  $\tau_1(x_1) - \tau_2(x_1)$ at $N$ uniformly spaced points in $D_{x_1}$ to identify regions of monotonicity. For each region, the sign of $\tau_1(x_1) - \tau_2(x_1)$ is checked at the edges of the region, and if the signs differ, the bisection algorithm is then used obtain a root. If no roots are found (which could occur, for example, from the presence of noise), $\tau_1(x_1) - \tau_2(x_1)$ is resampled in the region of the local minimum, and if still no roots are found, the local minimum is returned instead. Basic pseudocode for the sampling isoline recovery algorithm is provided in Algorithm \ref{algo_1}, while pseudocode for the monotonic region division and bisection algorithms is developed in Appendix \ref{psuedo} in Algorithm \ref{algo_3} and Algorithm \ref{algo_4}. 

\begin{table}[]
\begin{tabular}{l}
\hline
\begin{tabular}[c]{@{}l@{}}\textbf{Algorithm 1:} 
Pseudocode for the sampling isoline method. Inputs $d_l$, $S_i(E)$, and $\vec{M}(E)$ are  used implicitly\\
in function evaluations. Resampling when no solutions are found is not represented to simplify presentation. \\
Pseudocode for the Newt($x,f(x)$), find\_region($x_{samp},\ f(x_{samp}))$, and bisect($f(x),\ a,\ b$) functions is given \\
in Appendix \ref{psuedo}.\end{tabular} \\ \hline
\begin{tabular}[c]{@{}l@{}}1: \ \textbf{Input} $g_i^{M},\ d_l,\ S_i(E),\ \vec{M}(E$),\ $x_1^{(min)},\ x_1^{(max)},\ (x_2)_0,\ N$\\ 
2: \ \textbf{Function} $\tau_1(x_1)$ $\leftarrow$ Newt($(x_2)_0$,\ $g^R_1(x_1,\ x) - g_1^{M}$) \\
3: \ \textbf{Function} $\tau_2(x_1)$ $\leftarrow$ Newt($(x_2)_0$,\ $g^R_2(x_1,\ x) - g_2^{M}$) \\
4: \ \textbf{Function}  $f(x)$ $\leftarrow$  $\tau_1(x)- \tau_2(x)$ \\
5: \ $x_{sampled}$ = make\_grid($x_1^{(min)},\ x_1^{(max)},\ N$) \\
6: \ \textbf{for} $i$ in length($x_{sampled}$) \\
7: \ \qquad $\Delta\tau_{sampled}[i] \leftarrow f(x_{sampled}[i])$  \\
8: \ mono\_regions, $\Delta\tau_{monos}$ $\leftarrow$ find\_regions($x_{sampled}$,\ $\Delta\tau_{sampled}$)\\
9: \ \textbf{for} $i$ in length(mono\_regions) \\
10: \qquad$x_{1,i}^{left}$,\ $x_{1,i}^{right}$ $\leftarrow$ mono\_regions[$i$]\\
11: \qquad $\Delta\tau_i^{(left)})$,\ $\Delta\tau_i^{(right)}$, $\leftarrow$  $\Delta\tau_{monos}[i]$ \\
12: \qquad \textbf{if} sign$(\Delta\tau_i^{(left)}) \neq$ sign$(\Delta\tau_i^{(right)}$)\\
13: \qquad \qquad [$x_1^{sol},\ \tau_1^{sol}$] $\leftarrow$ bisect($f(x),\ x_{1,i}^{(left)},\ x_{1,i}^{(right)} )$\\
14: \qquad \qquad Roots $\leftarrow$[$x_1^{sol},\ \tau_1^{sol}$]\\
15: \textbf{if} no solutions\\
16: \qquad $x_1^{approx}$ $\leftarrow$ $x_{sampled}$[\textbf{where}($|\Delta\tau_{sampled}| =$ \textbf{min}$(|\Delta\tau_{sampled}|)$)]\\
17: \qquad $\tau_1^{approx}$ $\leftarrow \tau_1(x_1^{approx})$ \\
18: \qquad Roots $\leftarrow$ [$x_1^{approx},\ \tau_1^{approx}$] \\
19: \textbf{return} Roots
\end{tabular} \\ \hline
\end{tabular}
\refstepcounter{table}
\label{algo_1}
\end{table}

\subsection{Phantoms and Simulation}
We use two realistic phantoms to model the application of these frameworks to two separate material triplets. First, a realistic breast phantom was generated using the FDA VICTRE software for an uncompressed breast consisting of glandular tissue, adipose, and a spiculated tumor region filled with iodine contrast (14.79 mg I/mL) \citep{badano_evaluation_2018}. This material triplet was chosen for its potential application in contrast-enhanced dedicated breast tomography \citep{sarno_dedicated_2015, kim_adaptive_2021}. Linear attenuation coefficients for iodine were obtained from those published by NIST, while the glandular tissue and adipose values were taken from the Penelope 2006 Monte Carlo software \citep{hubbell_tables_2004, salvat_penelope2006_2008}. 

To show the utility of our noise propagation framework, we calculate values of $\text{det}|J(\vec{x}_l)|$ over the set of material path lengths present in the breast phantom to ensure uniqueness and obtain a prudent choice of tube potentials for this phantom. The SpekPy package was used to simulate the spectrum for a given tube potential for a beam with a fluence of $2.2\times10^4$ photons per detector \citep{poludniowski_technical_2021}. Then, for the selected tube pairing, we compare the result of 500 empirical noise realizations to the analytic CNR, defining the signal region as the tumor with iodine contrast and the background region as the glandular tissue. In particular, noisy 512 view by 512 detector sinograms of the breast phantom were produced, where the standard deviation of the path length noise was taken to be the size of one pixel on the reconstruction grid. Newton's method was used for sinogram recovery. The VMI's for monoenergies from 30 keV to 90 keV were then reconstructed on a 512 $\times$ 512 pixel (0.014 cm pixels) grid using FBP and used to obtain an empirical CNR. The analytic CNR at each monoenergy was then obtained from equation \ref{cnr_eq} and equation \ref{s2_eq}. Lastly, for this tube pairing, the matrix $Q$ and the eigenvector with the smallest eigenvalue was also computed and compared to sets of $w$ obtainable using linear attenuation coefficients (i.e., $w=M_j(E)/||\vec{M}(E)||_2$). 

The second phantom studied was a realistic heart phantom with a calcified cardiac plaque generated from a labeled CT scan. In particular, two versions of the phantom were produced. The first preserved the anatomy of the heart, and consisted of uniform regions of water, iodine contrast, and bone in order to study non-unique solutions with consistent data (i.e., decomposing a three-material object into a three-material basis). The second consisted of adipose, muscle, tissue, bone, hydroxyapatite, water, and iodine contrast (i.e., equation \ref{mu_eq} with $n=7$) to study the presence of non-unique solutions with inconsistent data (i.e., decomposing an object with more than three materials into a three-material basis). The mass fractions of the cardiac calcification were set to be 43.3\% hydroxyapatite and 56.7\% tissue, with linear attenuation coefficients obtained using the XCOM tool \citep{berger_xcom-photon_2010}. Using a tube pairing of 70 kVp and 140 kVp, selected using $\text{det}|J(\vec{x}_l)|$ to ensure the presence of non-unique solutions, the Speky package was used to generate noiseless consistent sinograms, noiseless inconsistent sinograms, and noisy inconsistent sinograms for beams with a fluence of 2$\times10^5$ photons per detector. For all three cases, the modified isoline algorithm was then used for basis sinogram recovery under a bone, water, and iodine basis assumption. In practice, for each sinogram element, the values of $x_1^{(min)}$ and $x_1^{(max)}$ were set dynamically based on $d_l$. For the initial guess $(x_2)_0$ in Newton's method, the longest length of the second basis material in the phantom was used.

When multiple solutions are returned for a sinogram element, a strategy needs to be developed to choose between them. One way to view the presence of multiple solutions is as occurring as a result of the loss of information during the integration over energy in equation \ref{meas_eq}. In our case, information about the object that produced the data is known perfectly and can be used to select a solution. In particular, we use a root-mean square logarithmic error to compare $(\vec{x_l})_{recovered}\cdot \vec{M}(E)$ and $(\vec{x_l})_{true}\cdot \vec{M}(E)$ curves for each sinogram element. Once a single solution was selected for each sinogram element, bone, water, and iodine basis images were reconstructed using FBP. 

\section{Results}

\subsection{Breast Phantom}
\begin{figure}[t]
\begin{centering}
\includegraphics[width=0.95\textwidth]{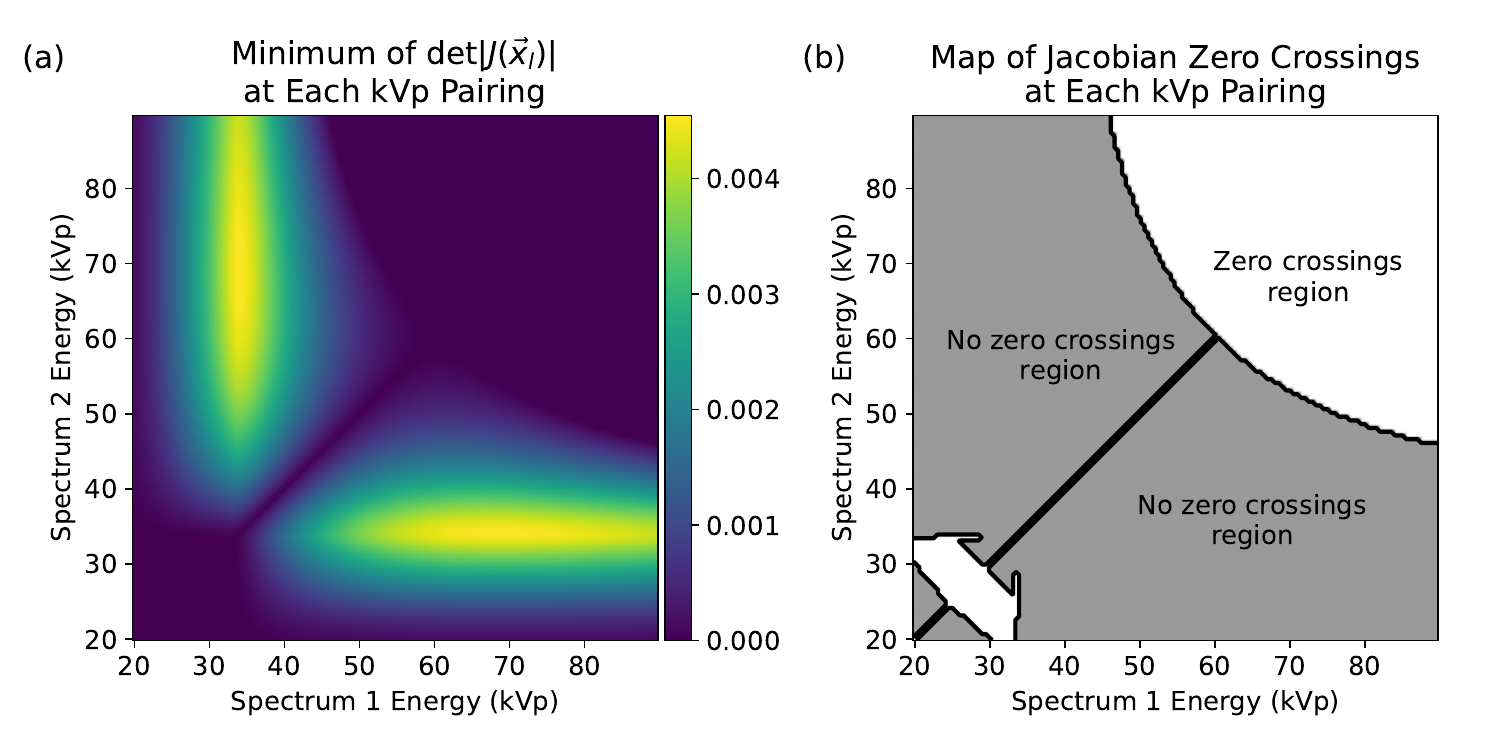}
\caption{Plots of $\text{min(det}|J(\vec{x}_l)|)$ (a) and logical map of the zero crossings of $\text{det}|J(\vec{x}_l)|$ over kVp pairings between 20 kVp and 90 kVp (b). Grey regions of (b) indicate $\text{det}|J(\vec{x}_l)|$ has no zero crossings for a kVp pair, while white regions indicate a zero crossing. The region of relatively high $\text{min(det}|J(\vec{x}_l)|)$ in the region of lower tube potentials from 30--40 kVp and higher tube potentials from 55--80 kVp indicates good noise properties for the sinogram decomposition, and the absence of zero crossings in the same region indicates uniqueness when performing sinogram decomposition.}
\label{jac_plot}
\end{centering}
\end{figure}

The realistic breast phantom contained a range of path lengths of adipose, glandular tissue, and iodine in the domain $\mathcal{V}=[0,\ 7.10]\times[0,\ 3.97]\times[0,\ 0.0007]$ (cm). Increasing the value of the determinant of $J(\vec{x}_l)$ reduces the material sinogram covariances and the mean pixel variance of the VMI. We chose the minimum value of $\text{det}|J(\vec{x}_l)|$ among all path lengths combinations $\vec{x}_l$ in $\mathcal{V}$ at a particular tube pairing as a figure of merit for that tube pairing because picking a tube potential where this value is high ensures a higher value of $\text{det}|J(\vec{x}_l)|$ for all sinogram elements. Figure \ref{jac_plot} shows the minimum of $\text{det}|J(\vec{x}_l)|$ at each tube pairing between 20 kVp and 90 kVp. Crucially, the minimum of $\text{det}|J(\vec{x}_l)|$ assumes a local maximum at 34 kVp and 69 kVp, and a region of relatively high minimum $\text{det}|J(\vec{x}_l)|$ values is evident for a lower tube potential of 30--40 kVp and a higher tube potential of 55--80 kVp. The Jacobian determinant also does not have any zero crossings in the region of kVp pairings with high minimum values of $\text{det}|J(\vec{x}_l)|$, so that the sinogram estimations should be unique in this region.

\begin{figure}[tbp]
\begin{centering}
\includegraphics[width=0.5\textwidth]{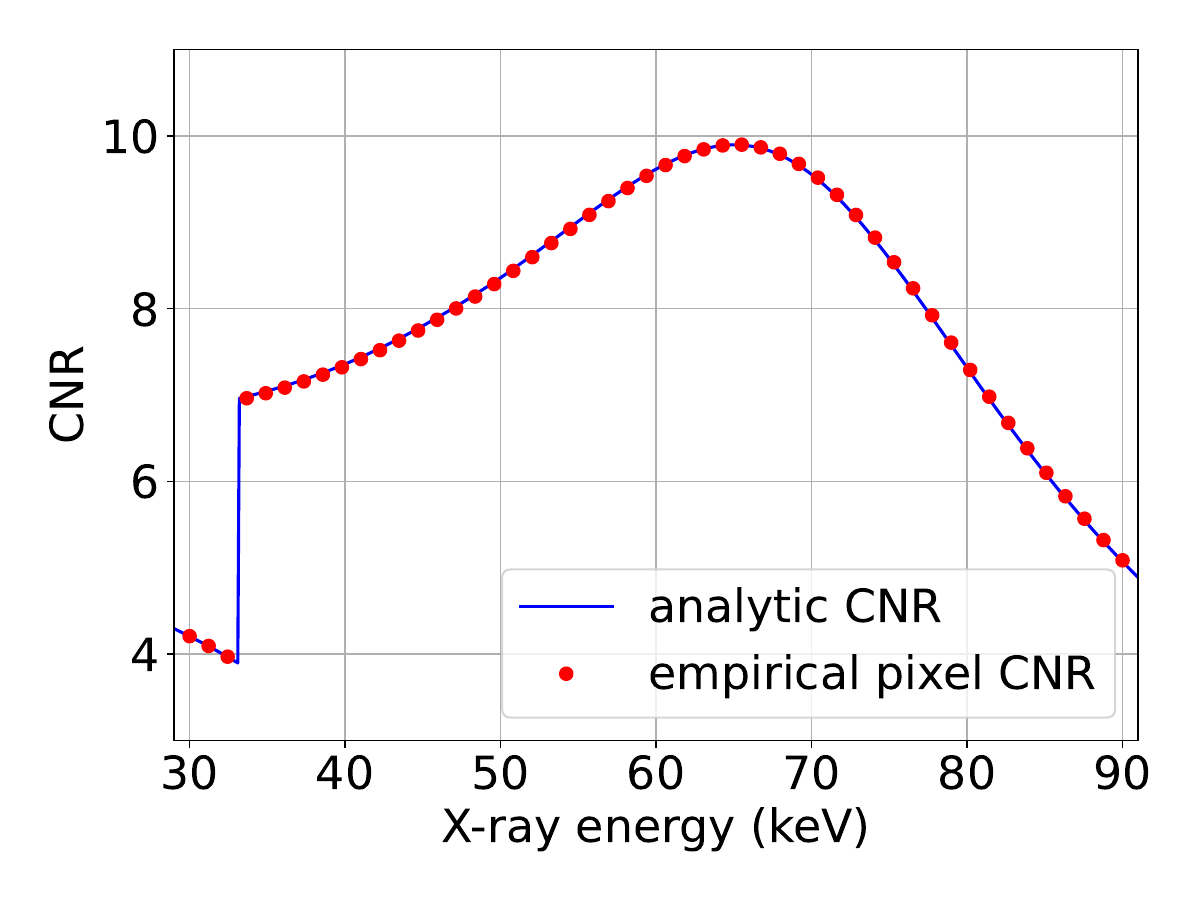}
\caption{Iodine CNR as a function of VMI monoenergy. The blue curve indicates the analytically predicted CNR calculated from equation \ref{cnr_eq} and equation \ref{s2_eq}. The red dots indicate the CNR calculated from 500 noise realizations, measured from the set of resulting VMIs. Note the sharp discontinuity near 33.2 keV which corresponds to the iodine k-edge.}
\label{cnr_plot}
\end{centering}
\end{figure}

Based on Figure \ref{jac_plot}, an ideal tube pairing of 34 kVp and 69 kVp was selected and used to produce 500 noise realizations. For each realization, VMI's were generated on a grid, which were then combined to calculate the CNR at monoenergies from 30 keV to 90 keV. Figure \ref{cnr_plot} shows the empirically realized CNR as a function of VMI energy plotted alongside the CNR predicted by our framework. The empirical and analytical curves show strong agreement. The CNR curve in this case exhibits a local maximum at 65.5 keV with a maximum CNR of 9.9. A large increase in iodine-glandular tissue contrast can also be seen moving across the iodine k-edge at 33.2 keV, caused by the increase in the linear attenuation coefficient of iodine and the weighting of its basis image in the VMI.

\begin{figure}[tbp]
\begin{centering}
\includegraphics[width=0.7\textwidth]{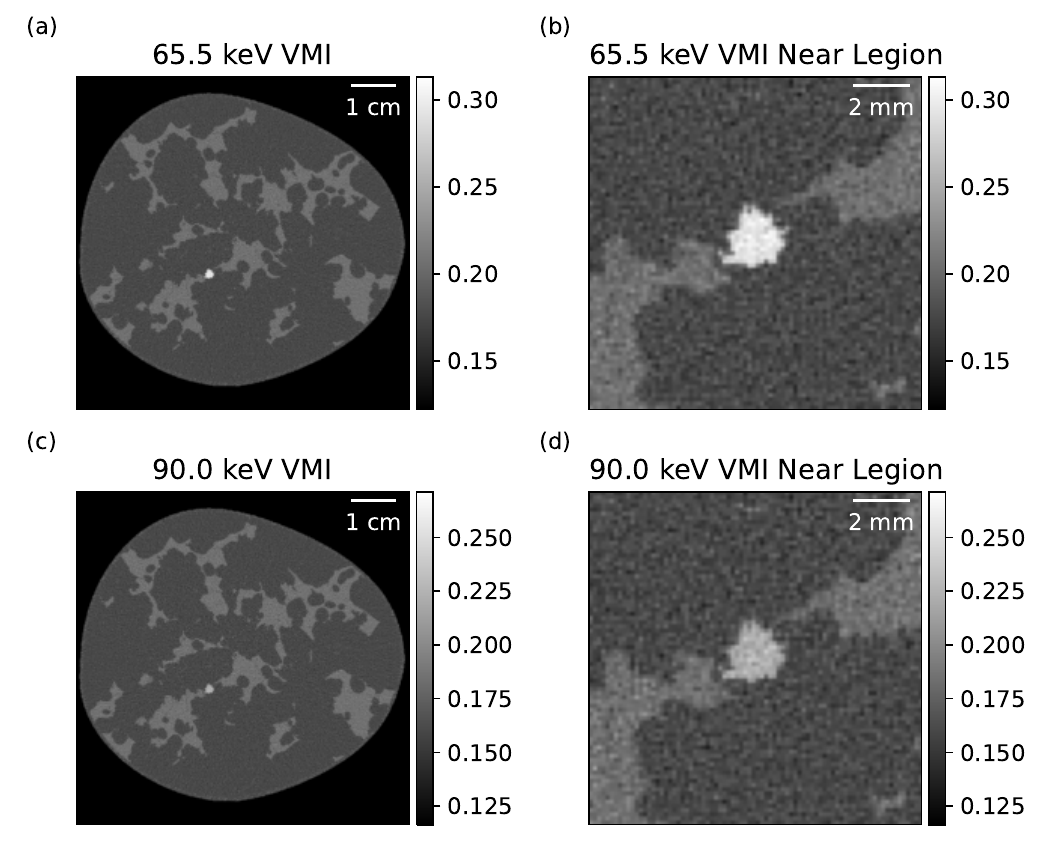}
\caption{VMI noise realizations of the realistic breast phantom at 65.5 keV (a) and 90.0 keV (c), and the reconstructions near the lesion at 65.5 keV (b) and 90.0 keV (d). Note that the grayscale is set so that adipose and glandular tissues appear at the same gray level.}
\label{vmis}
\end{centering}
\end{figure}

To highlight the change in CNR for a good and poor choice of VMI monoenergy, we present noisy VMI images of the breast phantom at 65.5 keV and 90.0 keV VMI energies in Figure \ref{vmis}, with the grayscale set such that iodine and glandular tissue are at the same grey level. The iodinated region of the 65.5 keV VMI has visually higher contrast relative to the 90.0 keV image, relative to fairly comparable noise backgrounds. It should also be noted that the CNR definition in equation \ref{cnr_eq} characterizes the pixel noise, that is, the average statistical fluctuations of each pixel in the iodine signal region. However, the noise in a particular noise realization, like those shown in Figure \ref{vmis}, is a combination of the structured reconstruction noise (i.e., variation over a uniform region that is not in the original phantom introduced by the reconstruction) and this statistical noise. The former standard deviation scales with the size of the iodinated region, while the latter does not, as it quantifies the noise of individual pixels. Additionally, the pixel variances calculated from many noise realizations and then averaged over the signal region (i.e., $s$) will not be the same as the variance assessed over the pixels in the signal region at a particular noise realization and then averaged over all the noise realizations, so that the ``CNR'' one would assess in an image visually as in Figure \ref{vmis} is not precisely the same CNR as defined in equation \ref{cnr_eq}. Both notions of the CNR should be related, so improving one (in our case by reducing random pixel fluctuations) should improve the other, as evident by the visual difference in contrast relative to the noise in Figure \ref{vmis}. 

The smallest eigenvalue of the matrix $Q$ was found to be $7.30\times10^{-8}$, with corresponding eigenvector $\vec{w}_{opt}=(9.9998\times10^{-1},\ 3.5583\times10^{-3},\ 4.6077\times10^{-3})$. For an ABI formed with this $\vec{w}_{opt}$, which should minimize the noise in the iodine signal region of the ABI, a CNR of 10.2 was calculated analytically.

\subsection{Cardiac Phantom}
\begin{figure}[tbp]
\begin{centering}
\includegraphics[width=0.45\textwidth]{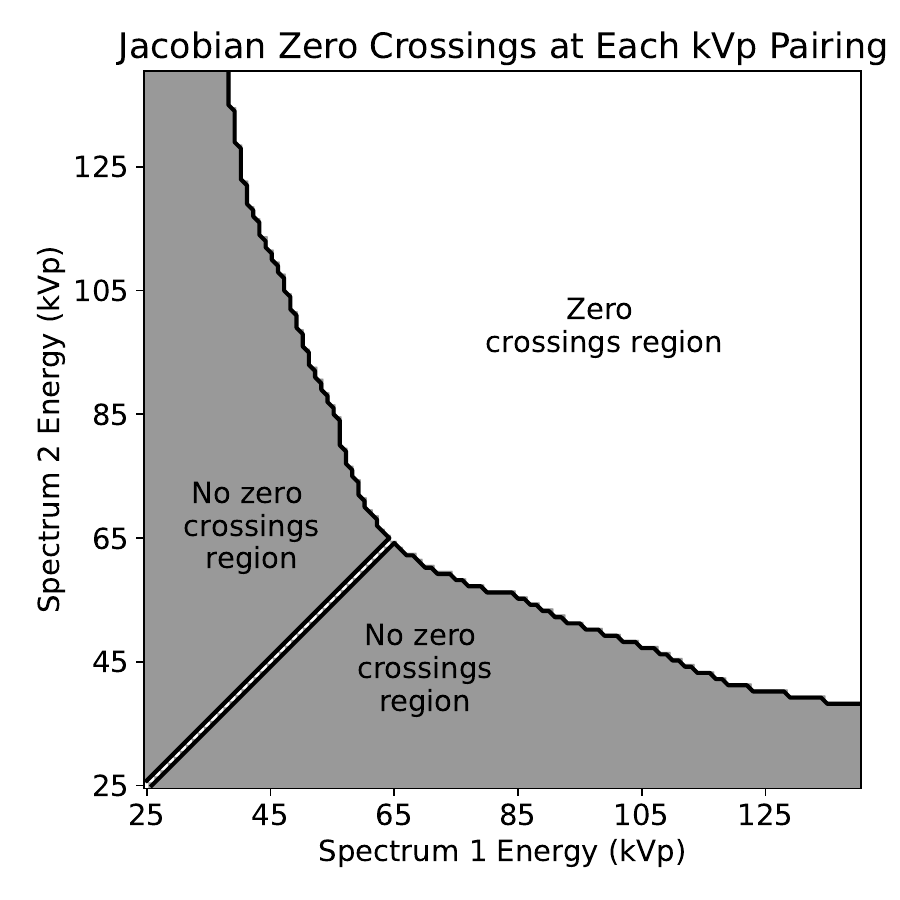}
\caption{Logical map of the zero crossings of $\text{det}|J(\vec{x}_l)|$ for tube pairings between 25 kVp and 140 kVp. Grey regions indicate areas where $\text{det}|J(\vec{x}_l)|$ has no zero crossings, while white regions indicate areas where $\text{det}|J(\vec{x}_l)|$ has a zero crossing, and uniqueness during sinogram decomposition is not guaranteed.}
\label{card_jac}
\end{centering}
\end{figure}

Over the path lengths of water, iodine, and bone present in the simplified three-material phantom (domain $\mathcal{V}=[0,\ 35.0]\times[0,\ 0.04]\times[0,\ 3.0]$ cm), the Jacobian determinant at a tube pairing of 70 kVp and 140 kVp was found to have a zero crossing, so that unique basis sinogram decompositions are not guaranteed. Figure \ref{card_jac} shows a logical map of where $\text{det}|J(\vec{x}_l)|$ has a zero crossing for tube parings between 25 kVp and 140 kVp. It can be seen that a large patch of tube pairings result in a sinogram recovery problem that is likewise poorly posed. 

\begin{figure}[tbp]
\includegraphics[width=1.0\textwidth]{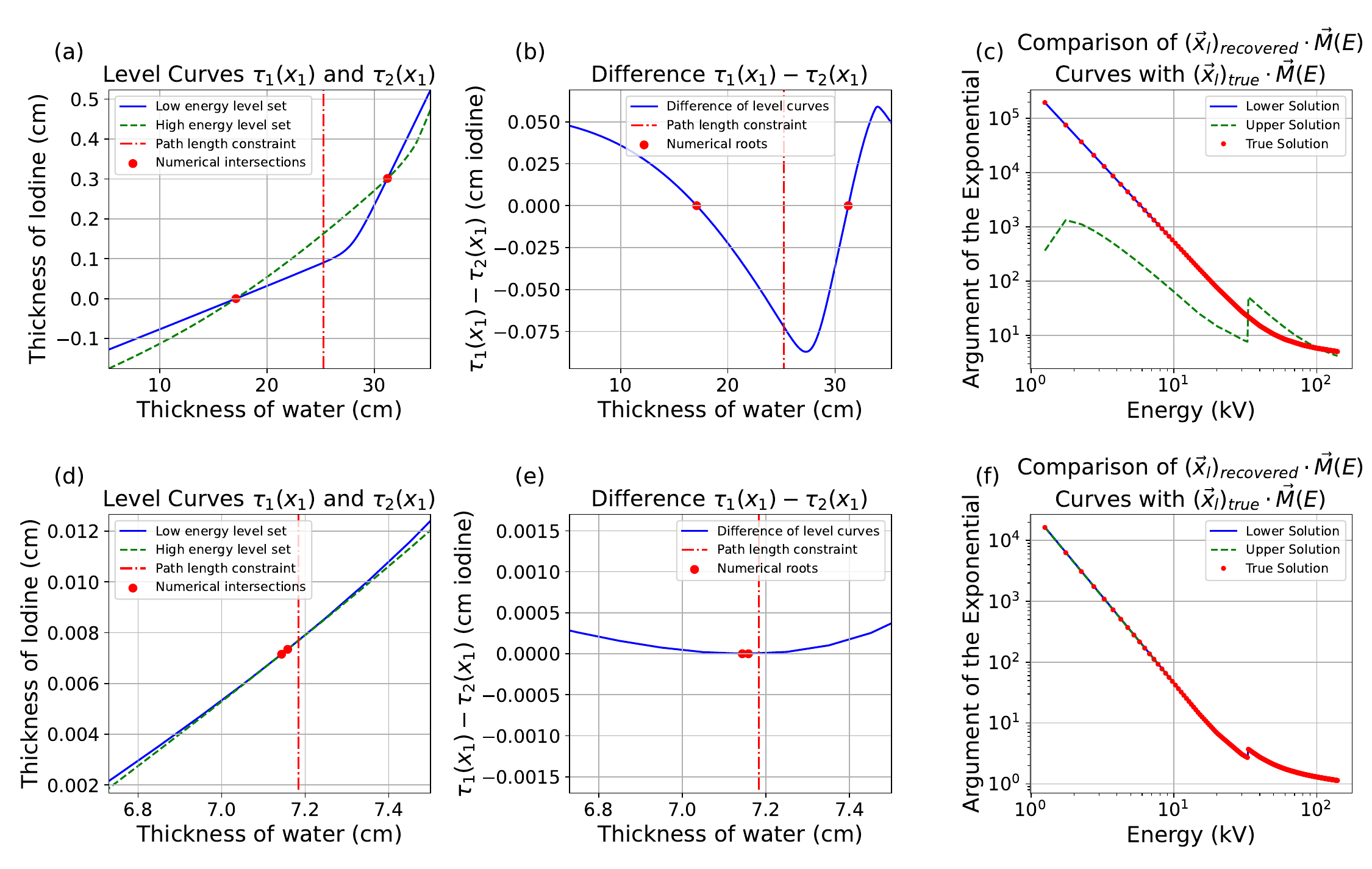}
\caption{Level curves for a single sinogram pixel in the consistent, noiseless phantom when either the lower solution (a) or the upper solution (d) is correct, their respective difference as used for root-finding (b and e), and $(\vec{x_l})_{recovered}\cdot \vec{M}(E)$ curves alongside $(\vec{x_l})_{true}\cdot \vec{M}(E)$ for each to illustrate how a root is selected (c and f). In both cases, the isoline curves clearly intersect twice, indicating non-unique solutions with just three materials present in the object, while the curves in the right column, generated using the two sets of roots found, demonstrate how a physical solution can be selected.}
\label{card_ls_con}
\end{figure}

\begin{figure}[tbp]
\begin{centering}
\includegraphics[width=0.82\textwidth]{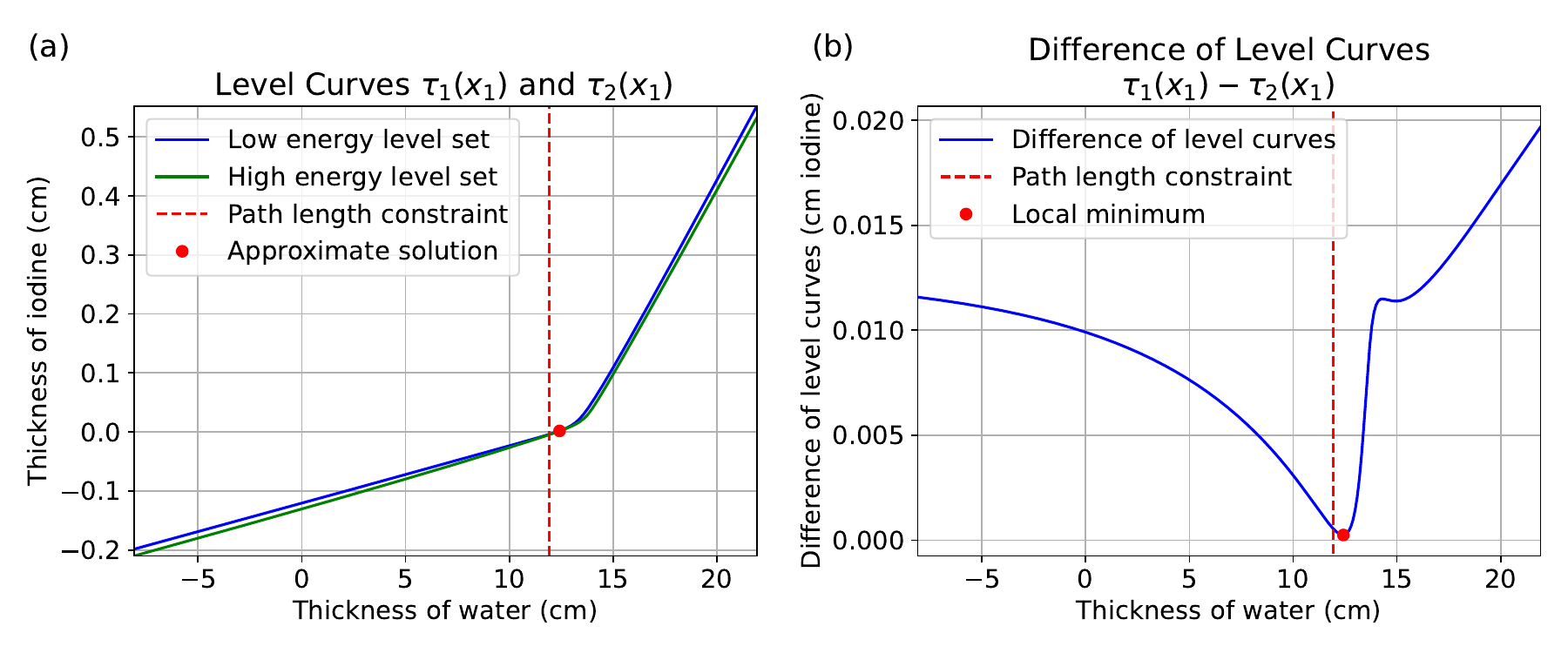}
\caption{Level curves for a single sinogram pixel in the inconsistent, noiseless phantom with no solutions (a), and their difference as used for root-finding (b). Here, when many materials are present in the sample, even in the absence of noise, the high and low energy level curves do not intersect, indicating there is no solution to the material decomposition for this sinogram element.}
\label{card_ls_incon}
\end{centering}
\end{figure}

For the consistent phantom, SIRA returned two roots for 99.7\% of sinogram elements (and no roots for the remaining elements), with differences in the recovered path length of water as large as 14.13 cm. Likewise, for the inconsistent phantom with no noise, SIRA returned two roots for 94.7\% of the sinogram elements, while with noise, 94.5\% of sinogram elements are returned with two solutions. Figure \ref{card_ls_con} shows the level curves $\tau_1(x_1)$ and $\tau_2(x_1)$ for the consistent phantom, their difference used for root-finding, and the $(\vec{x_l})_{recovered}\cdot \vec{M}(E)$ and $(\vec{x_l})_{true}\cdot \vec{M}(E)$ curves used for selecting a solution for a sinogram element where the lower solution is correct and for an element where the upper solution is correct. Notably, because the data is perfectly consistent, one of the two $(\vec{x_l})_{recovered}\cdot \vec{M}(E)$ curves and the $(\vec{x_l})_{true}\cdot \vec{M}(E)$ curve will match perfectly, while in the inconsistent case there will be some discrepancy between them. Then for the sinogram element represented in Figure \ref{card_ls_con}a-c, the solution with the lower path length of water should clearly be selected. Of course, this might also be determined by observing that the solution with the higher water path length in this case has a water path length that greatly exceeds to the maximum path length constraint, which is compensated by a negative path length of bone. Note that negative path lengths are generally allowed to avoid biasing solutions against noise, and so may not always form a good criterion for selecting solutions. It is should also be emphasized that both solutions may be physical in the sense that path lengths are non-negative and satisfy the path length constraint as shown in Figure \ref{card_ls_con}d-f, where the upper path length solution is correct. Figure \ref{card_ls_incon} additionally shows the level curves $\tau_1(x_1)$ and $\tau_2(x_1)$ and their difference used for root-finding for the inconsistent phantom in a case where no solutions are found. 

\begin{figure}[tbp]
\includegraphics[width=0.95\textwidth]{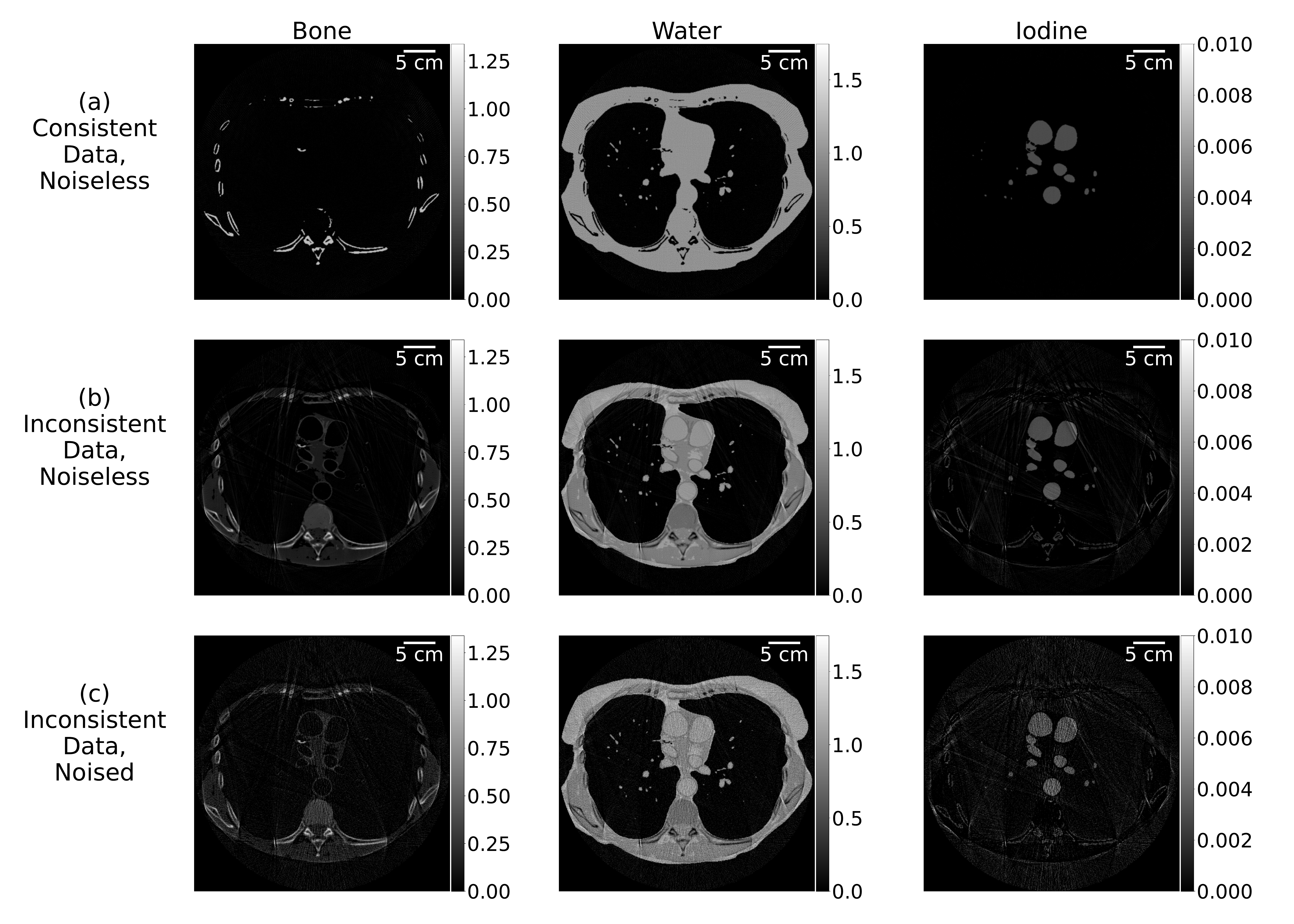}
\caption{Bone, water and iodine basis image reconstructions for the cardiac phantom, under a consistent choice of materials with no noise (a), an inconsistent choice of materials with no noise (b), and an inconsistent choice of materials with noise (c). The consistent noiseless case shows excellent material separation, while adding additional materials in the inconsistent case results in streak artifacts and some bony structures appearing in the iodine decomposition. Introducing noise adds a grainy texture to the reconstruction and some additional streak artifacts.}
\label{card_recon}
\end{figure}

The reconstructed basis material images produced from the recovered sinograms for the consistent phantom, inconsistent phantom, and inconsistent phantom with noise are shown in Figure \ref{card_recon}. The consistent case shows excellent separation between the three materials, while for both inconsistent data reconstructions, in addition to streak artifacts, some of the bony structures appear in the iodine basis image. In all three bone basis images, the cardiac plaque is clearly resolved.

\begin{figure}[tbp]
\includegraphics[width=0.95\textwidth]{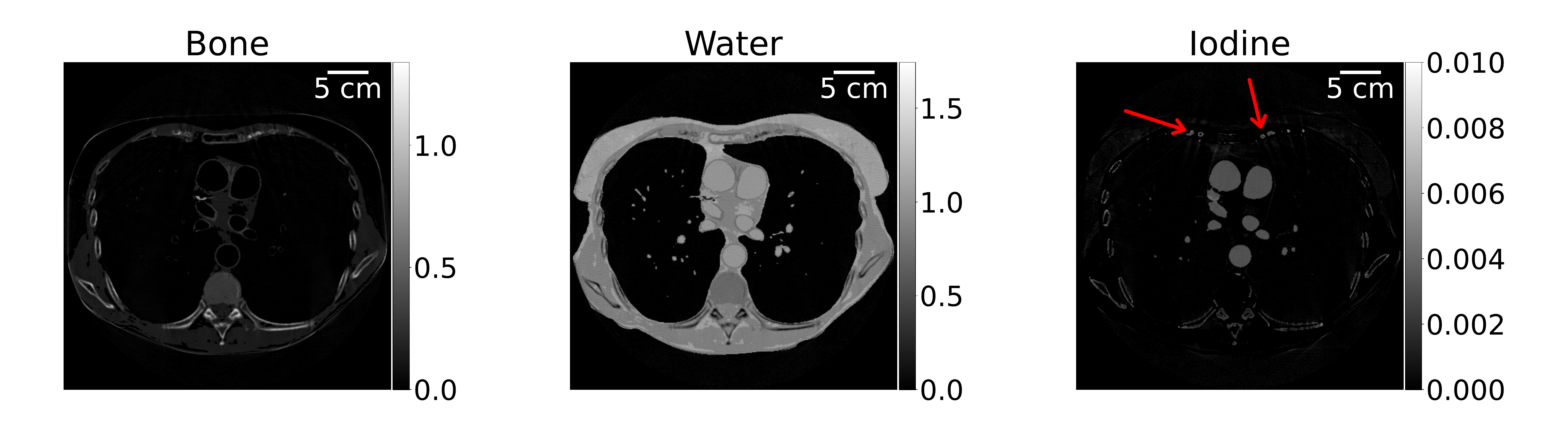}
\caption{Bone, water and iodine basis image reconstructions for the cardiac phantom, under an inconsistent choice of materials with no noise when the solution with the smallest path length of water is always chosen. While streak artifacts are reduced relative to Figure \protect\ref{card_recon}b, the appearance of bony structures in the iodine image is stronger, indicated by red arrows.}
\label{card_recon_left}
\end{figure}

For additional insight into the impact of choosing a solution when multiple are found, in Figure \ref{card_recon_left} we also show the resulting basis image reconstructions of the noiseless inconsistent phantom when the solution with the smallest path length of water is always chosen. Notably, always choosing the lower water path length solution appears to largely eliminate the streak artifacts seen in Figure \ref{card_recon}b when using $(\vec{x_l})_{recovered}\cdot \vec{M}(E)$ curves to pick the most physical solution. However, relative to Figure \ref{card_recon}b, the iodine basis image in Figure \ref{card_recon_left} has stronger contamination from the bone basis image, and some of structures in the bone basis image of Figure \ref{card_recon_left} appear with lesser contrast. 

 \FloatBarrier

\section{Discussion}
In this work, we studied the behavior of noise and the presence of non-unique solutions in DECT with a path length constraint. We used a noise propagation framework to select a tube pairing and VMI energy to optimize iodine CNR. For an ideal tube pairing where injectivity was confirmed over the material path lengths of interest in a realistic breast phantom, we were able to predict VMI CNR as a function of monoenergy and show strong agreement with the VMI CNR produced through simulating noise realizations. We demonstrated for the first time the presence of non-uniqueness in three-material DECT in a realistic chest phantom with iodine contrast and developed the isoline sampling algorithm to find multiple basis sinogram solutions when they occur.

The behavior of noise in VMIs produced from DECT measurements with data constraints in the image domain has been explored by \citet{ren_impact_2019}. Our approach differs by providing a framework to improve CNR by tackling the problem directly in the projection domain. Indeed, despite a variety of three-material decomposition DECT algorithms available \citep{liu_model-based_2022, long_multi-material_2014, mendonca_flexible_2014, liu_quantitative_2009}, the behavior of VMI CNR in these systems is not often studied outright. Moreover, the CNR in VMIs or ABIs may exhibit new behavior because when selecting $\vec{w}$ to form VMIs from three or more basis material images, there are significantly more possibilities relative to the two-material case.

For example, in the two-material case, $\vec{w}(E)$ effectively traces out some arc of a circle ($w_j=M_j(E)/||\vec{M}(E)||_2$) whereas the set of all possible $\vec{w}$ forms the complete circle. \citet{terzioglu_optimizing_2024} noted that this choice might be somewhat restrictive and leave open the possibility of a better CNR from a more general choice of $\vec{w}$, but found in practice for DECT, for a nice choice of basis materials, $\vec{w}_{opt}$ was contained in $\vec{w}(E)$ in all practical cases. For any given three-material basis, $\vec{w}(E)$ parameterizes a single curve constrained to the surface of a unit sphere, where all possible choices of $\vec{w}$ form the surface of the sphere. There are many choices of $\vec{w}$ that do not lie on this curve, and there is no guarantee that a $\vec{w}(E)$ that is used to construct a particular VMI will realize an image with a particular maximized image quality metric. Indeed, for the realistic VICTRE phantom, the best CNR with a strict VMI was 9.9, while a more arbitrary choice of weights in an ABI allowed for a mildly improved CNR of 10.2.

Using SIRA, reasonable basis image reconstructions of the consistent phantom, inconsistent phantom, and inconsistent phantom with noise were obtained. Notably, the majority of sinogram elements were found to have two solutions (at a minimum, 94.5\%), while the remaining sinogram elements had no solutions. Some cases where no solutions are returned are clearly a result of the algorithm sampling data on a discrete grid, when the roots are extremely close together. This happens, for example, in the strictly three-material case where 0.3\% of the sinogram elements return no solutions, but this cannot be true because there is at least one set of path lengths that satisfy the system of equations by construction. Failure to find a solution in this instance occurs if a sampling point is not selected between the two roots. Empirically, this was seen very near a local minimum, so that returning the local minimum is often sufficient for a satisfactory material decomposition and basis image reconstruction (see Figure \ref{card_recon}a). 

However, there are two notable instances where the algorithm returns no solution because none exists, where resolving the $\tau_1(x_1) - \tau_2(x_1)$ curve in Figure \ref{card_ls_incon} on increasingly fine grids does not return a root. The first instance occurs in the presence of noise, where a noisy measurement may shift the isoline curves so that they no longer intersect, as observed in the two-material case by \cite{phillips_dual-energy_2025}. The second instance occurs even without noise when working with the inconsistent data. This might occur because only a finite, primarily physical domain $D_{x_1}$ is searched for intersections and highly non-physical solutions exist outside this domain, requiring large path lengths of one material and large negative path lengths of another. Alternatively, in a basic sense, true no solution cases might also be expected because the data are no longer perfectly consistent and the three-material assumption has been violated. Considering that the dimensionality of the LAC space can be up to four, the set of three basis functions chosen might simply not span the given $\mu(E)$ approximated in equation \ref{mu_eq}. 

At first, the presence of no solutions resulting from insufficiently spanning LAC space seems contradictory with finding two solution cases, as one possible cause of non-unique solutions might be a system that is in fact over-described by the set of basis materials. For example, for an object consisting of entirely soft tissues with an intrinsic LAC dimensionality of two, it is easy to imagine that in a three-material basis (say water, adipose, and ICRU muscle), that non-trivial linear combinations of the LAC basis functions could effectively represent the same $\mu(E)$ function in equation \ref{mu_eq}. This would make such a three-material decomposition poorly posed from the start. Importantly, as shown in Figure \ref{card_ls_con}c, the presence of multiple solutions is clearly not caused by linear dependence among the basis material LACs, where the two linear combinations corresponding to unique solutions give clearly distinct LAC curves. Instead, non-unique solutions arise because the integral of these (exponentiated) curves with a weighting factor determined by the X-ray spectra result in the same value (see equations \ref{reduced_sys_eq1} and \ref{reduced_sys_eq2}). Thus, it is possible to have both instances where no solutions and multiple solutions could occur in the same phantom. 

One important limitation of this work is that the impact of basis material choice was not directly studied. Namely, while we do investigate two separate material triplets, work by \citet{tang_conditioning_2021} has shown that the choice of material basis can impact the condition number of basis functions and the behavior of VMI CNR as a function of monoenergy for material decompositions in photon counting CT. Similarly, the choice of basis functions in equation \ref{mu_eq} need not be limited to material LACs. In particular, \citet{weaver_attenuation_1985} demonstrate how principal component analysis can be utilized to generate alternative, non-physical or virtual basis functions, which can improve the stability of the representation. Given the potential of basis function choice to impact both noise and the condition number of basis functions, and the relatively broad choice of basis functions available, exploring different choices of basis function may prove fruitful. For example, it is possible that the poorly behaved difference in level set curves $\tau_1(x_1) - \tau_2(x_1)$ in Figure \ref{card_ls_con} and Figure \ref{card_ls_incon} might improve under an ideal material basis choice, making root-finding more reliable and improving the reconstruction. 

It is also worth commenting on the choice between multiple solutions during the sinogram recovery step with SIRA. Because this is a simulation work, we can use the starting basis material path length information to always select the most physical solution, giving rise to essentially perfect recovered basis images in the consistent data case. More generally, a strategy for picking between solutions must be developed, a limitation also encountered by \citet{phillips_dual-energy_2025} when presenting the isoline recovery algorithm. For example, one could always pick the solution with the smallest water path length, as we do in Figure \ref{card_recon_left}. For this phantom, always picking the solution with the lower water path length appears to give a reconstruction with less streak artifacts relative to the reconstruction picking the more physical solution. One possible cause for these streaks might be that in the inconsistent data case, mixing the set of smaller water path length solutions and the set of larger water path length solutions introduces strong discontinuities into the estimated sinograms, producing streaks. On the other hand, the smaller water path length solution only basis images in Figure \ref{card_recon_left} appear to have stronger bone contamination in the iodine basis image relative to the corresponding iodine basis image in Figure \ref{card_recon} where the physical solution is chosen.

\section{Conclusion}

In this work, we studied the noise properties of VMIs and the presence of non-unique sinogram decomposition in a three-material DECT system with a path length constraint. First, building on the work of \citet{terzioglu_optimizing_2024}, we derived analytic expressions for VMI CNR as a function of monoenergy for this system and used spectral theorem to find the weight $\vec{w}$ that gives the highest CNR when linearly combining basis images. Applying this framework to a realistic breast phantom, we identify an ideal dual energy tube pairing of 34 kVp and 69 kVp, for which we found an ideal VMI energy of 65.5 keV, as well as an arbitrary image combination with superior CNR with respect to the best possible VMI. Second, for a realistic chest phantom we studied det$|J(\vec{x}_l)|$ to find instances where injectivity was not guaranteed and demonstrated cases where the sinogram decomposition had multiple and no solutions. We then developed the sampling isoline recovery algorithm to find multiple solutions when they exist and showed reconstructions for the chest phantom composed of three materials with no noise, seven materials with no noise, and seven materials with noise. A major limitation of this approach and direction for future work was the use of a priori knowledge in selecting between solutions when multiple are found, as this is not available in any practical context so that a more general way of selecting solutions is needed. 

\section{Acknowledgments}
This research was supported in part by the National Institute of Biomedical Imaging and Bioengineering of the National Institutes of Health under award number R01CA287302, NIH S10 Award OD025081-01, and by the previous awards NIH S10 RR021039 and P30 CA14599. The contents of this paper are solely the responsibility of the authors and do not necessarily represent the official views of any of the supporting organizations.

 \FloatBarrier
\appendix
\section{Isoline Algorithm Subroutines}
\label{psuedo}
We describe three sub-routines utilized by the isoline algorithm, one for evaluation of $\tau_i(x_1)$, one for monotonic region division, and one for bisection. First, consider further the implicit evaluation of $\tau_i(x_1)$. Recall that $\tau_i(x_1)$ is a function that returns the value of $x_2$ that satisfies $g_i^R(x_1,\ x_2) - g_i^M = 0$. Then it is possible to evaluate $\tau_i(x_1)$ for a particular $x_1$ using a root-finding algorithm. In particular, we use Newton's method, which effectively fits a line at some starting guess, updates the guess using the root of the fitted line, and iterates until a desired tolerance is met. For the root-finding problem $f(x)=0$, and current root $x_n$, Newton's method gives the updated root as 
\begin{equation}
x_{n+1} = x_{n} - \frac{f(x_{n})}{f'(x_{n})}.
\end{equation}
Pseudocode for this general Newton's method is provided in Algorithm \ref{algo_2}. 
\begin{table}[]
\begin{tabular}{l}
\hline
\begin{tabular}[c]{@{}l@{}}\textbf{Algorithm 2:} Pseudocode for Newton's method evaluation of $f(x)$. 
\\The local variable $tol$ is the tolerance, which should be a value close 
\\to zero. In SIRA, the inputs $d_l$, $S_i(E)$, $D(E),$ and $\vec{M}(E)$ are used\\
implicitly in defining $f(x)$. 
\end{tabular} \\ \hline
\begin{tabular}[c]{@{}l@{}}1: \textbf{Function} Newton($x_0$, $f(x)$) \\ 
2: $tol$ $\leftarrow$ $10^{-8}$ \\
3: $i$ $\leftarrow$ 0 \\
4: \textbf{while}  $|f(x_i)| > tol $ \\
6. \qquad $x_{i+1}$ $\leftarrow$ $x_{i}- f(x_{i})/f'(x_{i})$ \\
7. \qquad $i$ $\leftarrow$ $i+1$ \\ 
8: \textbf{return} $x_i$ 
\end{tabular} \\ \hline
\end{tabular}
\refstepcounter{table}
\label{algo_2}
\end{table}

Specifically, for evaluating the implicit function $\tau_i(x_1)$ for a given $x_1^{const}$, the function $f(x)$ is just $g_i^R(x_1^{const},\ x) - g_i^M$. To compute the derivative of this expression, it is informative to construct the quantities $\phi^R_i$ and $\mathcal{I}^R_{i,j}$ (cf. equations \ref{int_eq}, \ref{i_note}, and \ref{deriv_eq}), where 
\begin{align}
	\phi^R_i(x_{l,1},\ x_{l,2}) &= \int_0^{V_1} S_1(E)D(E)A_l(E)e^{-M'_1(E)x_{l,1} - M'_2(E)x_{l,2}}dE \\
	\mathcal{I}^R_{i,j}(x_{l,1},\ x_{l,2}) &= \int_0^{V_1} S_1(E)D(E)A_l(E)M'_j(E)e^{-M'_1(E)x_{l,1} - M'_2(E)x_{l,2}}dE, 
\end{align}
recalling the notation of $M_j'(E)$ and $A_l(E)$ from equations \ref{reduced_sys_eq1} and \ref{reduced_sys_eq2}. Then the derivative of $f(x)$ can be obtained using 
\begin{equation}
\frac{\partial g_i^R(x_1,\ x_2)}{\partial x_{j}} = \frac{ \mathcal{I}^R_{i,j}(x_1,\ x_2)}   {\phi^R_i(x_1,\ x_2)},\ i,\ j=1,\ 2,
\end{equation}
and $n$th update step in Newton's method for evaluating $\tau_i(x_1)$ can be written as
\begin{equation}
(x_2)_{n+1}  =  (x_2)_{n} -  f((x_2)_n)  \cdot  \frac{\phi^R_i(x_1^{const},\ (x_2)_n)} { \mathcal{I}^R_{i,2}(x_1^{const},\ (x_2)_n) }.
\end{equation}
By iterating a sufficient number of times, a value of $x_2$ that satisfies $g_i^R(x_1^{const},\ x_2) - g_i^M = 0$ can be obtained. Note that this evaluation uses spectral information ($S_i(E)$), total path length information ($d_l$), and the energy-dependent linear attenuation coefficients ($\vec{M}(E)$). 

Given a means of evaluation $\tau_1(x_1)-\tau_2(x_1)$, to find roots of this quantity on the domain $D_{x_1}$, SIRA first requires dividing $D_{x_1}$ into regions where $\tau_1(x_1)-\tau_2(x_1)$ is monotonic. This is done by computing $\tau_1(x_1)-\tau_2(x_1)$ on a discrete grid of values in $D_{x_1}$. The difference in the value of $\tau_1(x_1)-\tau_2(x_1)$ between neighboring points is then calculated, and the signs of these differences can be used to find regions where  $\tau_1(x_1)-\tau_2(x_1)$ is always increasing or decreasing. Basic pseudocode describing this process is provided in Algorithm \ref{algo_3}, which takes a function evaluated on a grid as an input, and returns the monotonic regions of that function.

\begin{table}[]
\begin{tabular}{l}
\hline
\begin{tabular}[c]{@{}l@{}}\textbf{Algorithm 3:} Pseudocode for finding the monotonic regions of a function on\\
a discrete grid.  \end{tabular} \\ \hline
\begin{tabular}[c]{@{}l@{}}1: \textbf{Function} find\_regions($x_{sampled}$, $\Delta\tau(x_{sampled})$) \\ 
2: diffs $\leftarrow$ element-wise differences of the $\Delta\tau(x_{sampled})$ values \\
3: signs = sign(diffs) \\
4: sign\_changes $\leftarrow$ element-wise differences of signs \\
5: regions  $\leftarrow$  $x_{sampled}$[\textbf{where} sign\_changes = 0]  \\
6: $\Delta\tau_{in\_region}$  $\leftarrow$  $\Delta\tau$[\textbf{where} sign\_changes = 0]  \\
7: \textbf{Return} regions, $\Delta\tau_{in\_region}$ \\
\end{tabular} \\ \hline
\end{tabular}
\refstepcounter{table}
\label{algo_3}
\end{table}

Lastly, given a region of $\tau_1(x_1)-\tau_2(x_1)$ that is guaranteed to have a root (a monotonic region with a change in sign at the edges), we use the bisection algorithm to find the root. Bisection in particular is used because it is guaranteed to remain within the starting interval. Given a function $f(x)$ and two values bounding the root $a$ and $b$, the bisection algorithm evaluates $f(x)$ at the average $x=(a+b)/2$, and compares this to $f(a)$ to chose new bounding values. This procedure can be repeated until a root with the desired precision is obtained. Psuedocode for the bisection algorithm is provided in Algorithm \ref{algo_4}. 
\begin{table}[]
\begin{tabular}{l}
\hline
\begin{tabular}[c]{@{}l@{}}\textbf{Algorithm 4:} Pseudocode for bisection root-finding algorithm for a \\
single input function $f(x)$ given a starting interval [a,b]. Note the desired\\
tolerance $tol$ is set as a local variable. \end{tabular} \\ \hline
\begin{tabular}[c]{@{}l@{}}1: \textbf{Function} bisection($f(x)$, $a$, $b$) \\ 
2: $tol \leftarrow 10^{-8}$ \\
3: \textbf{while}  $|a -b| > tol $ \\
4: \qquad $c \leftarrow (a+2)/2$ \\
5: \qquad  \textbf{if} $f(a)\times f(c)<0$ \\
6. \qquad  \qquad $b\leftarrow c$ \\
7: \qquad  \textbf{else}  \\
8. \qquad  \qquad $a\leftarrow c$  \\
9: \textbf{return} $c$
\end{tabular} \\ \hline
\end{tabular}
\refstepcounter{table}
\label{algo_4}
\end{table}

\newpage
\bibliography{three_mat_bib_3.bib}
\bibliographystyle{apalike}

\end{document}